# Valuation of patents in emerging economies: a renewal model-based study of Indian patents

Mohd Shadab Danish[1], Pritam Ranjan[2], and Ruchi Sharma[1]

[1]Indian Institute of Technology, Indore
[2]Indian Institute of Management, Indore

**Abstract**

This study uses patent renewal information to estimate private value of patents by technology and ownership status. Patent value refers to the economic reward that the inventor extracts from the patent by making, using or selling an invention. Thus, we measure the value of patent right (private value of patent) from the patentee perspective. Our empirical analysis comprises of 555 patents with application year during 1999 to 2002. The term of these patents either ended in 2018 or lapsed due to non-payment of renewal fee. We model renewal decision of patentee as ordered probit where patent renewal fee increases with the age of patent. Variables such as patent family size, technological scope, number of inventors and grant lag are used as explanatory variables in the corresponding regression. Hence, this paper combines the patentee's renewal decision along with patents' characteristics and renewal cost schedule to estimate the initial rent distribution. We find that a large number of patents expire at an early stage leaving few patents with high value corroborating the results of studies using European, American and Chinese data. As expected, certain technology class patents enjoy high valuation.

Contact: MOHD SHADAB DANISH*, PRITAM RANJAN†, and RUCHI SHARMA‡, *Assistant Professor in Economics, Dr BR Ambedkar School of Economics University, Bengaluru, India (Email: shadab@base.ac.in), †Professor in OM&QT, Indian Institute of Management Indore, India (Email: pritamr@iimidr.ac.in), ‡Professor in Economics, Indian Institute of Technology Indore, India (Email: ruchi@iiti.ac.in)

## 1. Introduction

The rise of knowledge economy suggests that intellectual property rights particularly patents allows firms' to gain advantage vis-à-vis competitors. Patent system is intended to stimulate innovation giving exclusive right to patent owners for a limited period of time. Patents need to be valued, for a large set of transactions and for broader set of stakeholders to understand the technological strength of a country (Kamiyama et al. 2006). The increasing literature on different aspects of patent value indicates the need and relevance of capturing monetary and tangible value of patents.

In this paper, the concept of value of patent is defined as the private value of patent[i]. Value of patent is a multidimensional concept viewed from legal, economic and financial perspectives. Legal dimension of patent value refers to patent's sustainability when it is challenged (Burke and Reitzing, 2007). The financial accounting perspective is about the process of incorporating a patent in the financial statement of the company. The accounting valuation of patents is defined under various accounting standard related to intangible assets such as Accounting Standard No. 38 (IAS 38), International Financial Reporting Standards (IFRS) and US accounting principle. Economic patent value refers to the reward that a patentee is able to generate from a patent by excluding the competitors, licensing the technology to the third parties or a combination of both (Munari and Sobrero, 2011). Arora and Fosfuri (2003) defined patent rent as an incremental value above the profit captured without patent protection. The structure and inspiration of innovation are different in developing economies and therefore patent value in such countries may vary as compared to developed countries. This study intends to bring the discussion on patent valuation in India in the mainstream scholarship while highlighting the specific aspects of such valuation for an emerging economy.

The study focuses on value of residential patents assigned to firms'. Residential patents are categorized on the basis of their ownership status i.e. domestic and foreign, as defined by CMIE PROWESS. We also estimate patent values for different technology group to visualize the technology market in the country (e.g., Schankerman 1998; Deng, 2007). This study includes wide range of patent characteristics such as the number of inventors, grant lag, technology scope and family size, in the estimation process around Bessen's (2008) model.

The remainder of the paper is organized as follows. Section 2 starts with a previous literature on patent value and valuation methodology. Section 3 explains the methodology, the patent valuation model and parameters estimation techniques in detail. Section 4 elaborates the data and variables used for analysis. Section 5 summarizes descriptive statistics, and section 6 presents main findings. It also discusses uncertainty analysis and compares our results with other available studies in this field. Section 7 concludes the results.

## 2. Previous Literature
*2.1 Patent Value*

It is important to distinguish the value of underlying technology that a patent protects from the value of patent per se. Arora et al., (2001) define patent value as 'patent premium', referring to the extra value that a patent generates for the firm. Schankerman and Pakes (1985) estimate the private value of patents considering patent life as an indicator of its value. The underlying rationale using patent renewal approach is that the patent holder will not renew its patent for an additional year if the cost of holding it exceeds the revenue generated.

The private value of patent in China is measured by Gupeng and Xiangdong (2012) using renewal payment based information. The study compares the values of domestic and foreign patents (U.S., Japan and European countries) and finds that the value of Chinese patents is much lower than the value of foreign patents. On the similar line, Zhang et al. (2014) found that 30 percent patents filed by Chinese firms are renewed over four years, whilst this ratio is much higher for the U.S., Japanese and E.U. patents (between 40 % and 60 %) indicating the technology gap between China and developed countries. Similarly, Liu et al. (2014) found similar results in the case of Chinese agricultural patents. Entity wise private firms are more likely to renew their patent than public entities. Furthermore, agricultural biotechnology and agricultural chemicals patents are found with higher value than others.

In Indian context, there are a number of studies that utilize patent and R&D information to estimate various economic indicators (Kanwar and Hall, 2015; Ambrammal and Sharma, 2015). Kanwar and Hall (2015) estimate market value of R&D in the context of manufacturing firms in India. Ambrammal and Sharma (2015) utilize patent count information to measure its impact on firm performance. However, there is no study that explores the valuation aspect of Indian patent. Thus, the present study attempts to bring patent value discussion in Indian academia. By using Indian patents information from the economic point of view.

*2.2 Patent Value Estimation*

The estimation approach of monetary value of patents is classified into three broader groups. The first approach rely on the observed behavior of patent owners, for example analyzing patent renewal decision or assessing the economic terms of actual patent licensing (Bessen, 2008; Gupeng and Xiangdong, 2012). The second approach is based on the survey method in which investors are directly asked to provide an estimate of the value of their patents (Scherer and Harhoff 2000). The third approach is based on the valuation made by external investors, either by stock market valuation or by venture capital valuations (Hall and MacGarvie 2006).

Earlier studies on patent valuation have used number of patents and owners' characteristics including number of inventors, co-application and size of applicant, patent citations received, and characteristics of the owner to determine the initial return of the patent. In the same spirit, we estimate profit flows as a function of a diverse set of information about the patent such as family size, number of inventors, technological scope and grant lag. Overall, this study contributes to the literature in the following ways. First, no study before this has used full length renewal information of the patent to measure the average life of the patent in India by technology differences. Second, no study has quantified patent system incentive in India in monetary terms. Third, earlier study used Monte Carlo simulation techniques (see Bessen, 2008) to estimate private value of patents; however, this study uses evolutionary techniques of simulation (GAs) which gives more robust estimates of parameters[ii]. Thus the obtained result using diverse set of factors give an edge over previous studies. To check the robustness of the result we conducted additional sensitivity analysis. In the next section, we will discuss the methodology for which we closely follow the approach proposed in Bessen (2008).

### 3. Methodology

Model proposed by (Pakes and Schankerman 1984) is based on the life of a patent in which the patentee decides to keep the patent in force to internalize the streaming returns. For every granted patent, there is a compulsory renewal fee if the patentee wants to keep it enforced. The sequence of renewal fees increases monotonically with age and is denoted by $c_{it}$. Patentee who pays renewal fee earns implicit return $r_i(t)$, from the patent protection during the active life of the patent. We assume that $r_i(t)$ is known to the patentee at $t = 0$, the time of application / filing the patent. In a more complicated model, (Schankerman and Pakes 1985) allow patentee to be uncertain about return's sequence.

## 3.1. Renewal Model

We make two key assumptions about the profit flow of patent. First, the returns of patents, $r_i(t)$, depreciate at a fixed rate. Though stochastically varying depreciation rate based model may appear to be more flexible, (Bessen 2008) demonstrates that model based on the constant depreciation rate leading to similar results as the models using variable depreciation rate. That is,

$$r_i(t) = r_i(0)e^{-dt},$$

where d is a fixed (unknown) depreciation rate, and $r_i(0)$ is the initial return at the time of application / filing the patent. The annual renewal fee $c_{it}$ is also assumed to depend only on time t and not the patent characteristics. Following (Bessen 2008), we further modelled the present value of profits from t to $t + T$ (here, $T = 1$ annual renewal cycle) as,

$$r_i(0)z_t = \int_t^{t+1} r_i(\tau)e^{-s(\tau-t)}\, d\tau,$$

where,

$$z_t = e^{-dt}\left(\frac{1 - e^{-(d+s)}}{d + s}\right),$$

The discount rate s is different from the technological depreciation or decay rate d. Following (Bessen 2008), we assume that an expected return to patent value depreciates at a constant rate. In this study, s is fixed at 0.1[iii].

The second key assumption is that the initial return is lognormally distributed. Let $X_i$ denote the vector of characteristics for the i-th patent. Then,

$$\ln(r_i(0)) = \boldsymbol{\beta} \cdot X_i + \varepsilon_i, \tag{1}$$

where $\varepsilon_i$ is independent and identically distributed (iid) normal variables with mean zero and (unknown) variance $\sigma^2$. To model the initial returns of a patent, this study uses four patent characteristics which include size of the family size, inventor size, technology scope and grant lag and technology class dummies in the equation. The most crucial part of the renewal model is to formulate the decision criterion for deciding whether or not a patent should be renewed at time t. The necessary and sufficient condition for a renewal of the i-th patent at time t is

$$\ln(r_i(0)) \geq \ln\left(\frac{c_{it}}{Z_t}\right).$$

Let $T_i$ be the expiry age of the i-th patent. Based on the data we have, the i-th patent will fall into one of the following three scenarios at time $2 \leq t \leq 19$:

(a) The patent is never renewed: $[T_i = 2]$. The i-th patent is never renewed if and only if the value of the patent at the end of the second year is less than the renewal cost, i.e.,

$$\log(r_i(0)) \leq \log\left(\frac{c_{i2}}{Z_2}\right).$$

Following the log-normal distribution, the probability of this event can be computed by

$$P[T_i = 2] = P\left[\log(r_i(0)) \leq \log\left(\frac{c_{i2}}{Z_2}\right)\right] = \Phi\left(\frac{\log\left(\frac{c_{i2}}{Z_2}\right) - \boldsymbol{\beta} \cdot X_i}{\sigma}\right), \quad (2)$$

where $\Phi$ is the standard normal cumulative distribution function (CDF).

(b) The i-th patent is renewed until maturity: $[T_i = 20]$. It is sufficient to say that this event can occur only if the i-th patent was renewed at $t = 19$, i.e.,

$$\log(r_i(0)) \geq \log\left(\frac{c_{i,19}}{Z_{19}}\right),$$

with probability

$$P[T_i = 20] = P\left[\log(r_i(0)) \geq \log\left(\frac{c_{i,19}}{Z_{19}}\right)\right] = 1 - \Phi\left(\frac{\log\left(\frac{c_{i,19}}{Z_{19}}\right) - \boldsymbol{\beta} \cdot X_i}{\sigma}\right) \quad (3)$$

(c) The i-th patent expires prematurely: $[3 \leq T_i \leq 19]$. In other words, the i-th patent expired at time $T_i = t$ and it was renewed at time $t - 1$, i.e.,

$$[T_i = t] = \left[\log(r_i(0)) \geq \log\left(\frac{c_{i,t-1}}{Z_{t-1}}\right)\right] \cap \left[\log(r_i(0)) \leq \log\left(\frac{c_{it}}{Z_t}\right)\right],$$

with probability

$$P[T_i = t] = P\left[\log\left(\frac{c_{i,t-1}}{Z_{t-1}}\right) \leq \log(r_i(0)) \leq \log\left(\frac{c_{it}}{Z_t}\right)\right]$$

$$= \Phi\left(\frac{\log\left(\frac{c_{it}}{Z_t}\right) - \boldsymbol{\beta} \cdot X_i}{\sigma}\right) - \Phi\left(\frac{\log\left(\frac{c_{i,t-1}}{Z_{t-1}}\right) - \boldsymbol{\beta} \cdot X_i}{\sigma}\right). \quad (4)$$

These probabilities are not computable as the model parameters $\Omega = (\sigma, d, \boldsymbol{\beta})$ are unknown. Thus, we have to use the data on expiry age $(T_i)$ and different characteristics $(X_i)$ to estimate the model parameters, which are then used to simulate the initial patent value $r_i(0)$.

## 3.2. Parameter Estimation

Assuming $\boldsymbol{\beta}$ is a 9-dimensional vector of regression coefficients; we have to estimate 11 parameters. We follow the maximum likelihood approach for estimating the model parameters $\Omega = (\sigma, d, \boldsymbol{\beta})$. For the data on n patents, the likelihood based on the distribution of $T_i$, presented in Equations (2) - (4), is given by

$$L(T_1, T_2, \ldots, T_n \,; \Omega) = \prod_{i=1}^{n} P(T_i = t_i). \qquad (5)$$

Unfortunately, none of parameter estimates can be found in a closed analytical form. Thus, a numerical optimization approach has to be used for estimating the parameters. We follow an evolutionary optimization technique called the Genetic Algorithm (GA) (Holland, 1975) for finding the maximum likelihood estimates (MLE) of $\Omega = (\sigma, d, \boldsymbol{\beta})$.

The search space for parameter are defined by $d \in (0.1, 0.5)$, $\sigma > 0$, $\beta_{grant-lag} < 0$, $\beta_{technology-scope} > 0$, $\beta_{family-size} > 0$, and $\beta_{inventor-size} > 0$, whereas the other $\beta$ coefficients were allowed to take any value in the real line[iv].. In this GA, we used the initial population size of 10,000 and 20 generations for estimating the parameters. Furthermore, we adopted the multi-start approach to reduce the dependency of the initial population and find robust estimates of $\Omega$.

The final estimates of $\Omega$ were taken as the median of the best 200 solutions from the last generation of the GA process (see annexure II). The standard errors of these 200 solutions were used to quantify the uncertainty and sensitivity of the parameter estimates.

## 3.3. Simulation of the Patent Values

Using the parameter estimates $(\hat{\sigma}, \hat{d}, \hat{\boldsymbol{\beta}})$ we estimate the bounds for each patent value conditional on corresponding renewal decisions made by patentee. Using Monte Carlo simulation we estimate the initial return $r_i(0)$ of the patent, thus $r_i(t)$ value is calculated using fixed depreciation rate as demonstrated in studies by (Bessen 2008; Maurseth 2005).

The bounds on $\varepsilon_i$ for the i-th patent conditional on the observed renewal decision can be deduced separately for the three cases as listed in Section 3.1.

(a) Patent expires at the end of the second year (i.e., patent is never renewed)

$$\varepsilon_i \leq \ln\left(\frac{c_{i2}}{Z_{2(\hat{d})}}\right) - \hat{\beta}X_i. \qquad (6)$$

(b) Patent expires prematurely (i.e., at t = 3,4, ... ,19),

$$\ln\left(\frac{c_{i,t-1}}{Z_{t-1(\hat{d})}}\right) - \hat{\beta}X_i \leq \varepsilon_i \leq \ln\left(\frac{c_{it}}{Z_{t(\hat{d})}}\right) - \hat{\beta}X_i. \qquad (7)$$

(c) Patent matures at 20$^{th}$ year from the date of filling

$$\ln\left(\frac{c_{i,19}}{Z_{19(\hat{d})}}\right) - \hat{\beta}X_i \leq \varepsilon_i. \qquad (8)$$

For every observation of the Monte Carlo iteration, we select $\varepsilon_i$ as a random draw from the log-normal distribution of Equation (1) determined by $\hat{\beta}$, $\hat{\sigma}$ and $\hat{d}$. The Monte Carlo simulations were repeated a large number of times ($10^6$) to ensure that we had sufficient number of observation for estimating each patent value. The estimates of $r_i(t) = r_i(0)e^{-\hat{d}t}$ can also be used to find the present value of all expired patents at time t,

$$V(T) = \sum_{t=1}^{T} r_i(t) - c_{it}(1+i)^{-t}, \qquad (9)$$

Where, $r_i(t)$ and $c_{it}$ denotes return and renewal cost of patent i at time t respectively. Whereas, s denotes annual discount rate which is fixed at 10%.

**4. Data and Variables**

*4.1. Data*

The study contains 1135 firms' patent data filed at Indian Patent Office (IPO) during January 1999 to December 2002 and the details of patent information were extracted from PATSEER database. Patents filed during 1999 will expire in 20 years that is 2019 and patents filed in year 2002 completes its 20 years in 2022 (maximum life of patent is 20 years). The renewal period in this study ranges during January 2001 to October 2018.

In the data cleaning process, we found that only 752 patents had complete information. Out of 752 patents, 26% are non-expired patents and were removed from the sample. Hence, the final sample of this study is contains 555 firms' granted patents. Gupeng and Xiangdong (2012) suggested that the expired patent based studies are more useful for accumulated but terminated resources up to the investigation date. Therefore, in this study we have taken only patents for which complete information was available.

Each patent is categorized into five technology class as per the IPC 2008 classification namely \chemical, electrical, mechanical, instruments and 'other field'. The technologies are classified on four digit IPC level (for example see Table 1 for technology level).

Table 1. IPC Classification-2008

| Section | Class | Sub class | Group | Sub group |
|---------|-------|-----------|-------|-----------|
| A | 61 | K | 31 | /545 |

Source: WIPO- IPC Technology concordance-2008

We calculated the average life of patent for cohorts (i.e. 1999, 2000…2002) separately in Table 2. The patents filed in 1999 have average length 9 years which increases to 10 years in 2000. Later in 2001 and 2002, it was stagnant at 8 years. Average patent length of Indian patent is 8.89 years during the sample period.

4.2. Renewal fee

As per Section 53, Rule 80 of Indian patent act 1970, every patent holder is required to pay patent maintenance fee annually (3rd year onwards from the date of application) after the grant to keep a patent in force. In this study, we follow the fee structure as per "The patents rules 2003"[v] of Indian patent act 1970. There has not been any change in the renewal fee schedule till 2018. Note that India and China have annual renewal fee payment requirement, which is different from the US (see Table 3).

Table 3. Patent renewal fee schedule for India, China and US

| Renewal Years | US | Renewal Years | China | Renewal Years | India |
|---------------|------|---------------|----------|---------------|---------|
|  |  | 1 to 3 | $135.00 | 3 to 6 | $54.81 |
| 4-7 | $1,600 | 4 to 6 | $180.00 | 7 to 10 | $164.43 |
| 8-11 | $3,600 | 7 to 9 | $300.00 | 11 to 15 | $328.86 |
| 12-14 | $7,400 | 10 to 12 | $600.00 | 16 to 20 | $548.10 |
|  |  | 13 to 15 | $900.00 |  |  |
|  |  | 16 to 20 | $1,200.00 |  |  |

Note: Renewal fee information is taken from respective patent office website (USPTO. SIPO and IPO)

Theoretical model by Baudry and Dumont (2006) establishes that an increase in patent renewal fee would proportionately discourage low-quality patents. Rassenfosse and Jaffe (2018) empirically find that an increase in the renewal fee led to the weeding out of low quality patents. Moore (2005) finds that a significant numbers of patents issued each year at USPTO expire before completing twenty years. Thus, it is clear that renewal fee create de facto differentiation in patent value. Similar to other developed and developing countries, India follows the incremental renewal fee to remove the worthless patents from the system. The reasons for non-payment of renewal fee could be appropriated to the fact that patent owner understand the economic idea of sunk cost and therefore reduce their losses by letting less valuable patent expire.

The value of patent can be revealed based on its owner's assessment of patent's cost and benefits. Many studies in past hypothesized that renewal fee create a recurring investment and therefore it is expensive for patent holder to keep a patent in force till its statutory life limit particularly in a situation where renewal fee is increasing in nature (Baudry and Dumont, 2006). However, the criticism of renewal model is that it measures patent value from the patentees' point of view. Further, such valuation excludes other incidental expenses such as attorney costs, company internal costs and therefore value of patents are likely to be underestimated (Pitkethly, 1997).

### *4.3. Regression Variables*

Patent document provides details about technical, legal and business specific aspects. These characteristics are likely to have a bearing on patent value. Hence, we quantify patent specific aspects to find their association with patent value. The following patent characteristics are used in this study.

Technology Scope: Lerner (1994) observes that technological breadth of patent is significantly associated with the firm's valuation, and broad patents are more valuable when many possible substitutes in the same product class are available.

Patent Family Size: The set of patents filed in several other countries which are close or related to each other by one or several other priority filing is referred to as family size. Lanjouw et al. (1998) found that number of jurisdictions in which the patent has been sought is associated with the value of patent.

Grant lag: It is defined as the time elapsed between the filing date of application and date of grant.

Number of inventors: Among others, Gupeng and Xiangdong (2012) used number of inventors' information as a determinant of economic value of patent.

To further explain whether variables in this study are correlated with each other or not we generate the correlation table. The correlation matrix is presented in Table 4 which shows no high correlations among the variables. The VIF values for Technology scope, Inventor size, Family size and Grant lag are 1.03188, 1.0394, 1.0515 and 1.0215, respectively, which supports the absence of multicollinearity among the predictors.

Table 4. Correlation matrix of regression variables

|  | **Grant Lag** | **Family size** | **Inventor Size** |
|---|---|---|---|
| **Family size** | -0.0247 |  |  |
| **Inventor Size** | 0.0753 | 0.1337 |  |
| **Technology scope** | -0.124 | 0.2399 | 0.0299 |

## 5. Descriptive Statistics

The sample consists of 555 patents granted to firms'. These patents are dis-aggregated into five technology groups and ownership status (Indian firms' and foreign subsidiary in India). Table 5 presents the summary statistics of independent regression variables for the data. The average grant lag for complete sample is 7.25 which is more than any developed nations average. The larger grant lag brings attention toward the patent system in India. This could be improved by introducing latest technology in the examination process and speeding up work with experts in the field.

Table 5. Technology category-wise summary statistics of the regression variables

| | | Chemical | Mechanical | Instruments | Electrical | Others field |
|---|---|---|---|---|---|---|
| | Obs. | 237 | 100 | 31 | 170 | 17 |
| Technology scope (Number of technological class) | Mean | 1.04 | 1.04 | 1.13 | 1.05 | 1.29 |
| | Std. Dev. | 0.23 | 0.19 | 0.56 | 0.23 | 0.68 |
| | Min | 1 | 1 | 1 | 1 | 1 |
| | Max | 3 | 2 | 4 | 2 | 3 |
| Inventor Size (Number of people named as an inventor) | Mean | 2.74 | 2.14 | 2.74 | 2.31 | 1.58 |
| | Std. Dev. | 1.74 | 1.45 | 1.89 | 1.78 | 1.27 |
| | Min | 1 | 1 | 1 | 1 | 1 |
| | Max | 9 | 8 | 8 | 12 | 6 |
| Family Size (Number of jurisdiction in which a patent is sought) | Mean | 4.49 | 0.69 | 6.61 | 2.44 | 2.41 |
| | Std. Dev. | 8.5 | 3.51 | 14.1 | 5.83 | 7.44 |
| | Min | 0 | 0 | 0 | 0 | 0 |
| | Max | 36 | 27 | 62 | 29 | 29 |
| Renewal Years (Number of years a patent is survived) | Mean | 8.62 | 10.1 | 10.1 | 8.54 | 7 |
| | Std. Dev. | 5.46 | 5.64 | 5.98 | 6.16 | 4.98 |
| | Min | 0 | 0 | 0 | 0 | 0 |
| | Max | 20 | 20 | 20 | 20 | 15 |
| Grant lag (Number of years elapsed between application and grant date) | Mean | 6.85 | 7.44 | 7.67 | 7.31 | 7 |
| | Std. Dev. | 1.8 | 1.56 | 1.75 | 2.04 | 2.23 |
| | Min | 2 | 4 | 4 | 2 | 5 |
| | Max | 14 | 12 | 11 | 16 | 13 |

Source: Authors' calculations.

The mean technological scope is higher in "others field" followed by instrument and electrical patents. Chemical and mechanical patents have equal level of technological scope. This implies that other filed patents are having wider claims. Average number of people involved in a patent is highest in the chemical and instrument fields whereas electrical and mechanical have slightly smaller inventor size. The average number of family size is highest in the instrument field (6.61) and it is lowest in the mechanical (0.69) field.

Table 6. Technology-wise patent expiration at different age (in percentage)

| Technology Category | Never Renewed | 3rd to 6th year | 7th to 10th year | 11th to 15th year | 16th to 20th year | Total Patents |
|---|---|---|---|---|---|---|
| Chemical | 23.207 | 4.641 | 30.380 | 36.709 | 5.063 | 237 |
| Mechanical | 17.000 | 4.000 | 27.000 | 37.000 | 15.000 | 100 |
| Instruments | 19.355 | 3.226 | 29.032 | 29.032 | 19.355 | 31 |
| Electrical | 28.824 | 1.176 | 27.059 | 34.118 | 8.824 | 170 |
| Others | 29.412 | 0.000 | 52.941 | 0.000 | 17.647 | 17 |
| Average | 23.559 | 2.609 | 33.282 | 27.372 | 13.178 | 555 |

Source: Authors' calculation.

Table 6 depicts survival rate of resident patents at different age. The renewal fee in India is relatively very small as compared to many developed nations. Contrary to the common understanding, a large number of patents (23.56%) are not renewed even for a small amount of renewal fees, and only 33.28% patents are maintained over the age of 10. This clearly indicates that a large number of patents are actually of lower value. The early expiry of patent in India could be appropriate to the low technology life cycle. In some areas technologies are fast changing and therefore patents associated with those technologies become irrelevant for owners.

Yi (2007) finds that average patent life is longer in Germany and the U.K compare to other countries such as Belgium and Austria. However, in Germany only 70 percent patents survive up to 10 years and about 50 percent of patents lapsed by age 14. The median length of patent life in Austria and Belgium is 11 years. The average renewal period of Chinese patents ranging from 3.29 to 5.94 years which is obviously shorter than U.S., Japanese and E.U. firms (4.31 to 9.06 years) (see Zhang et al., 2014). However, in India average patent length is 8.87 years which is higher than Chinese patents but lower than many technologically advanced countries.

## 6. Results and Discussion
### 6.1. Factors Influencing Patent Value

The estimated regression coefficients reported in this study accounts for the behavior of various characteristics (see Table 7). Please refer to equation (1) for settings of the underlying regression model. All parameter estimates are significant at 1% level of significance. These results confirm general findings about patents value indicators and its association with the patent value. A negative grant lag co-efficient indicates that a larger gap between the application date and grant date produces lessens the value of patent. Many studies in the past on patent valuation have mentioned about the negative association between grant lag and patent value (see Régibeau and Rockett, 2010). The reason could be appropriated to the fact that shorter commercial life of a patent with high grant lag is not able to generate much benefit to the inventors. However, the patented technology with longer commercial life will not be affected by long grant lag. However, during this "patent pending period" technology can be produced, sell and advertised by copier till the patent is issued. Therefore, higher grant lag eventually reduces the possibility of higher profit margin for the inventors (Hegde et al. 2016). The other side of the coin is that higher grant lag in some cases opens a big opportunity for the inventors when they can sue imitators and take over the market share of the infringing product. Since the imitator has already invested huge money on the product development they will have no option except buying the license from the original inventors. Such litigation disputes are rare in India and therefore longer grant lag inversely affect the patent value instead of generating any gain to the inventors.

Unlike USPTO, Indian patent office does not provide additional time onto the term of patent if the granting delayed more than three years from the filing date. In this study, we find that a significant number of patents lapse in initial years. Such patents share some common characteristics such as higher grant lag (7.45) and lower family size (2.99).

Table 7. Maximum likelihood estimates of $\Omega = (\beta, \sigma, d)$ obtained via Genetic Algorithm

| Technology Field (Reference category 'others') | $\beta_{techfield}$ (standard error) |
|---|---|
| Chemical# | -2.04(0.03) |
| Mechanical# | 1.81(0.06) |
| Electricals# | 2.45(0.07) |
| Instruments# | -0.40(0.04) |
| Patent Characteristics | $\beta_X$ (standard error) |
| Family size | 0.37(0.01) |
| Inventor size | 0.21(0.01) |
| Grant lag | -1.47(0.01) |
| Technology scope | 0.78(0.04) |
| $\sigma$ | 6.07 (0.03) |
| $d$ | 0.49(0.00) |
| Total Observation | 555 |

Note: All values in parenthesis are standard error. # denotes dummy variable. p-values for all estimates are less than $10^{-6}$, and hence significant at 1% level.

When compared in technology sectors with consumer appliances (that is 'other') electrical and mechanical patents value are positive whereas instrument patents and chemical which includes pharmaceuticals patents are negative. In India, law did not allow product patents for pharmaceutical sector during 1970-2005; hence, our sample consists of only process patents. Thus the results are not surprising because worldwide process patents have lesser value in comparison to the product patents.

We find the depreciation rate $d = 0.49$ which suggests that the expected value of patented technology in India depreciates at much higher rate than such a technology in China (24.28%, according to (Gupeng & Xiangdong 2012). While estimating, we found that technology depreciation rates cannot be fixed in 0.1 to 0.25 ranges as found in other studies. Our optimization results improved when we increased the upper boundaries of depreciation rate. Fast depletion or short technology time cycle could be the reason for higher depreciation rate in India.

### *6.2. Initial Returns by Different Technology Field and Ownership Group*

We now discuss the trend and pattern in the predicted $r_i(0)$ values. Table 8 reports the mean and median values of initial returns for different technology fields on 2001 base price.

On the line of previous research, we find that the value distribution of Indian patents is highly skewed. Thus, the number of patents used for the measure of innovation output is not a good measure. The private value of patents renewed for the 10th years is 34 times more valuable than those patents renewed only for 5 years. The patent renewed for complete 20 years are 273 time more valuable than 10 years old patents. Excluding the first 2 years (because renewal fee is not applicable), we find that patents renewed for an additional year signals that a patent is worth 2 times more valuable compared to patents not being renewed.

The result of this study suggests that the value distribution among different technologies is not similar. Since medicine sector was not open for product patents other sectors such as instruments, electrical and mechanical patents dominated the list. Here it is reported that patents of instrument technology have highest mean value whereas patents in chemical and "other field" have least mean value. Figure 2 presents the category-wise distribution of simulated patent values, $log(r_i(0))$. Irrespective of technology type, large number of Indian patents has lower value and only few block buster patents hold large value. Patents of chemical technology are highly skewed and electrical patents value are relatively least skewed.

Table 8. Estimated mean value ($M 2001) of initial returns by technology categories and ownership status [Mean of $r_i(0)$]

| Technology/Ownership | Patent Share | Value Share | Mean ($M) | Median ($M) |
|---|---|---|---|---|
| Chemical | 43.09 | 29.42 | 0.196 | 0.0341 |
| Mechanical | 17.53 | 21.39 | 0.350 | 0.0341 |
| Instruments | 5.41 | 7.75 | 0.411 | 0.0341 |
| Electrical | 30.78 | 40.60 | 0.378 | 0.0209 |
| Other field | 3.17 | 0.81 | 0.074 | 0.0209 |

| Foreign Subsidiary | 33.58 | 29.04 | 0.248 | 0.0341 |
| Indian Firms' | 66.41 | 70.95 | 0.306 | 0.0341 |

Note: Technology categories are from WIPO-technology classification (2008). All monetary values are in units of million U.S. dollars in year 2001 value.

Figure 1 shows the upward trend of initial return for the compete sample. Patents that are never renewed have more or less similar value. The value of patent shows upward with the increase in renewal age. The value differences among the patents can be observed in the trend line.

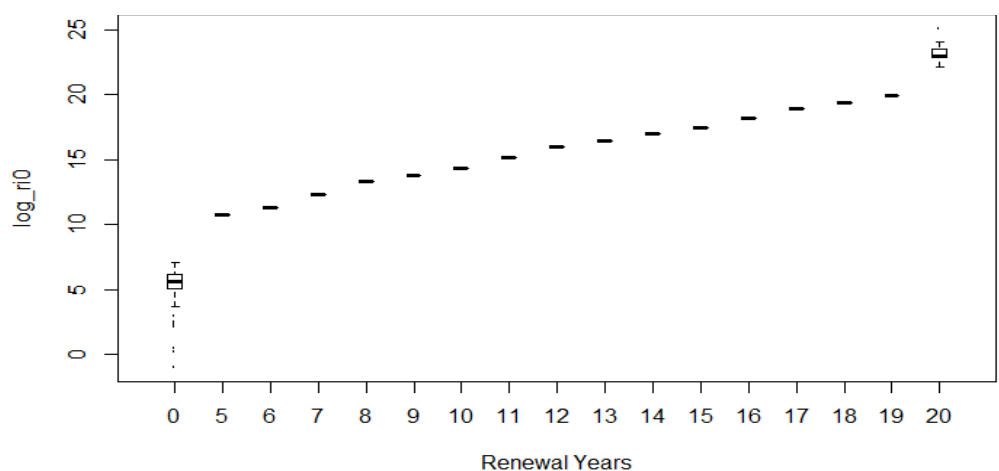

Figure 1. Distribution of estimated initial return $r_i(0)]$ of full sample with the age.

### 6.3. Estimation of Net Present Value

Net present value (NPV) of patent is estimated by discounting net returns at 10 percent discount rate to compare with earlier studies. All the values are calculated on 2001 price. Table 9 reports the mean and median of net present value (NPV) of patents at constant price (2001). The highest mean NPV is in the instrument field, followed by electrical, mechanical, chemical and 'others'. The highest median value is in the mechanical field followed by instruments, chemical, electrical and 'others'. As per the ownership category, patents belonging to foreign subsidiary in India have largest mean value as compared to Indian firms' patents. For easy comparison with European and US studies we have converted NPV to 2001 dollar.

The results obtained in this study give the lower bound of the private value of a granted patent in India. The lower bound private value of patent could be appropriated to several reasons. First, this study has only used patents assigned to India by IPO. Second, we have accounted only renewal costs, not application, drafting and attorney costs as does Putnam (1996). Therefore, the estimated value of patent doesn't reflect the total patent protection. Third, the cost of enforcement has not been included here as Lanjouw (1998), hence the estimated private value is again a lower bound of the real value. Fourth, we assume that renewal for each patent is independent from others and hence the strategic value of patent is ignored. Fifth, we do not allow for learning as Pakes (1984) discusses in the stochastic model of patent renewals. In a similar study, Grönqvist (2009) find that the value distribution of Finnish patents is skewed. At 25 percent quantile overall value of Finnish patent is reported to 326 Euro. On the other hand, our study on Indian patent observes 0.002 million dollar i.e. 2000 US dollar value of all patents at 25 percent quantile (see Table 11).

Table 9. Total Share of patents and mean Net Present Value ($M 2001) by technology and ownership status of patents assigned to India at IPO

| Technology/Ownership | Patent Share | Value Share | Mean ($M) | Median ($M) |
| --- | --- | --- | --- | --- |
| Chemical | 43.09 | 28.76 | 0.388 | 0.064 |
| Mechanical | 17.53 | 21.60 | 0.716 | 0.072 |
| Instruments | 5.41 | 7.88 | 0.847 | 0.072 |
| Electrical | 30.78 | 40.98 | 0.774 | 0.044 |
| Other | 3.17 | 0.76 | 0.140 | 0.042 |
| Foreign Subsidiary | 33.58 | 29.08 | 0.503 | 0.064 |
| Indian Firms' | 66.41 | 70.91 | 0.621 | 0.067 |

Note: Technology categories are from WIPO-technology classification (2008). All the values are reported in million U.S. dollars in year 2001.

Figure 2 presents the share of patent and its value by technology field and ownership status. The highest numbers of patents are coming from chemical and pharmaceutical sector but in value terms these sectors contribute less for reasons mentioned earlier.

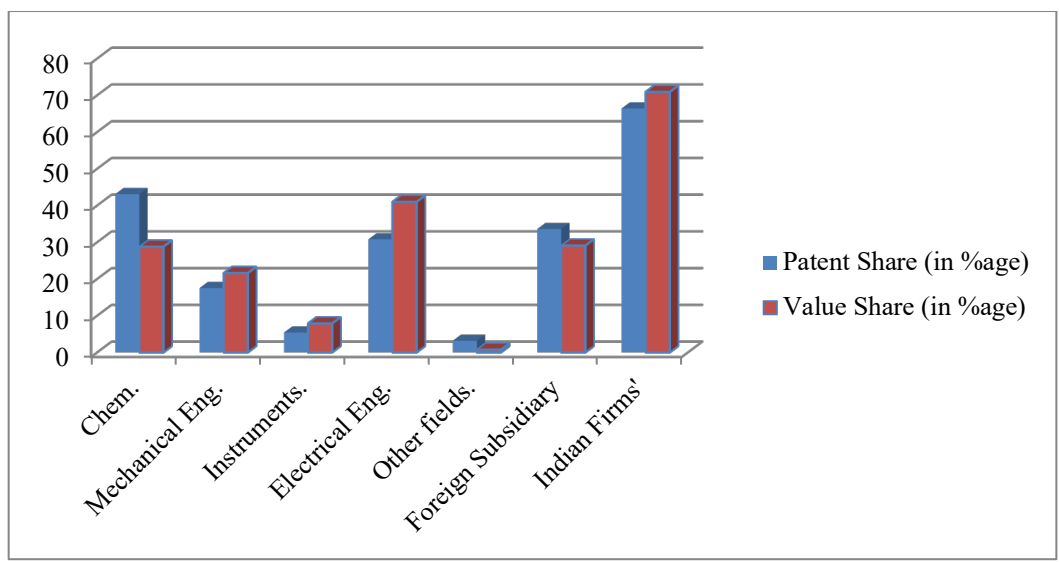

Figure 2. Share of patent value with percentage share of patents in different technologies and ownership

Estimated value of expired patents for different technology groups and ownership status, conditional on $\varepsilon_i$ from equation (a), (b) and (c) in Section 3.3 enables us to measure value of different set of patents based on the renewal fee in the local currency. Patent value of different owners and technology field are compared in tables 10 & 11. Table 10 presents the quantile distribution of patent value among different technology. The distribution of patents in left and right of median is extremely skewed.

Table 10. Distribution of the discounted lifetime value of patent right

| Quantile (%) | Chemical | Mechanical | Instruments | Electrical | Consumer appliances and Others |
|---|---|---|---|---|---|
| 25% | 0.002 | 0.010 | 0.010 | 0.000 | 0.000 |
| 50% | 0.064 | 0.072 | 0.072 | 0.044 | 0.042 |
| 75% | 0.357 | 0.631 | 0.583 | 0.583 | 0.072 |
| 90% | 1.488 | 1.619 | 3.282 | 1.619 | 0.357 |
| 95% | 1.554 | 3.405 | 3.486 | 3.376 | 1.488 |
| 99% | 3.373 | 7.472 | 7.452 | 12.200 | 1.488 |
| Mean | 0.388 | 0.716 | 0.846 | 0.774 | 0.140 |
| Std. Dev. | 1.002 | 1.494 | 1.642 | 2.172 | 0.358 |
| Obs. | 231 | 94 | 29 | 165 | 17 |

Note: Table 10 reports the simulated value distribution of Indian patents, for each technology field group. All monetary values are in units of million U.S. dollars in year 2001 value.

### 6.4. Uncertainty Analysis

Thus far, we have first estimated the parameters $\Omega = (\sigma, d, \boldsymbol{\beta})$ by taking the median value of the best 200 candidates from the Genetic Algorithm, and then used $\hat{\sigma}, \hat{d}$ and $\hat{\boldsymbol{\beta}}$ to determine $\log(r_i(0))$ for the i-th patent via Monte Carlo simulations. This approach does not account for the uncertainty in the parameter estimation process. As a result, we propose a slight modification in the estimation of $r_i(0)$.

The main idea is to use all 200 good $\Omega_k$'s (obtained from the final generation of GA) for predicting 200 realizations of $r_i(0)$ and then find the average (or median) of $r_i(0)$ as the predicted return for the i-th patent. That is,

$$\log(\hat{r}_i(0)) = \frac{1}{200} \sum_{k=1}^{200} \log\left(\hat{r}_i(0 \mid \hat{\Omega}_k)\right).$$

This approach not only leads to a more robust estimate of $r_i(0)$, but also yield the uncertainty estimate of the patent value prediction. Figure 8 presents 200 realizations of $r_i(0)$ for all patents considered in this paper. The patent value prediction for the median parameter estimates is also overlaid for reference. It appears that the predicted patent values over 200 realizations of $\widehat{\Omega}_k$ are very similar. It is important to emphasize however that this proposed approach will give more comprehensive interval estimates.

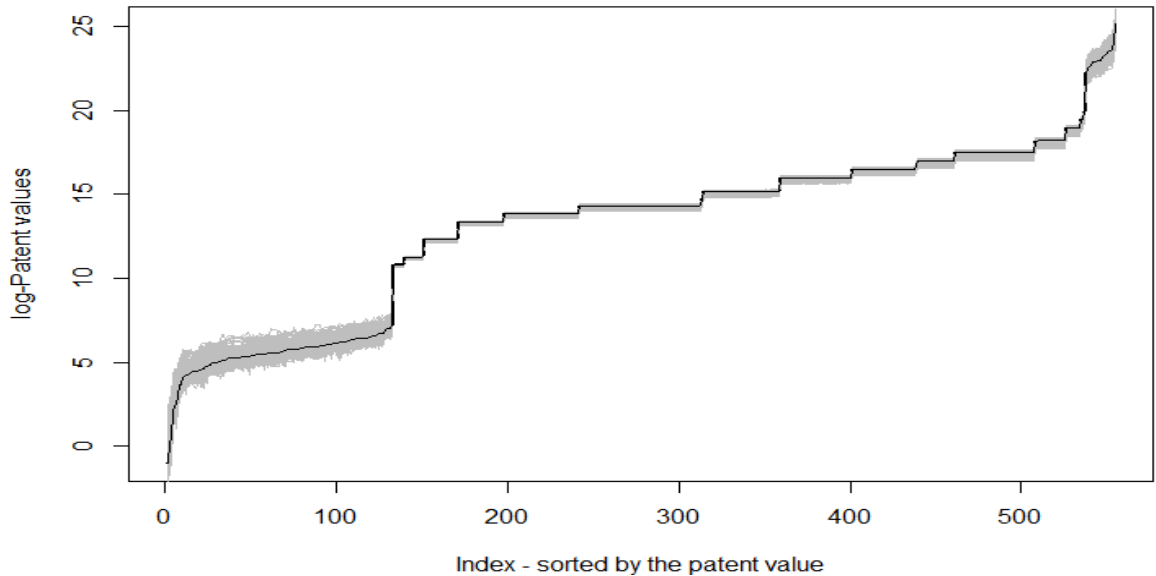

Figure 5. Sensitivity index line: Solid black line shows initial returns and the grey line represents predicted value over 200 realizations.

### 6.5. Comparative Analysis

This study enables us to compare our patent value with the estimates in the existing literature around the globe (see Table 12).

Mean net present value of foreign organization in US as reported by (Bessen 2008) is $2.905 million on $1992 price. Whereas, in India net present value of the foreign subsidiaries patents are reported $0.113 million dollar at $2001 price. The size of economy and quality of innovation are dissimilar between these two countries therefore, the differences in net present value are pretty much understandable.

### 7. Conclusion

Our empirical analysis suggests several implications with regard to patent valuation practices. The results provide an interesting finding about the Indian patents competitiveness across the technology. We also compare with other existing literature on the patent valuation (U.S and China). In line with other studies, we observe similar results of the patent and inventor characteristics (family size, technological scope, inventor size and grant lag) on initial return of the patent (Grönqvist, 2009). The average survival rate of Indian patents (8.37) is greater than Chinese patents (4.36). When compared to the technological advanced countries Indian patents life span is shorter. This implies that the quality of R&D in India or any developing nation for that matter is not sufficiently large. Therefore, the outcome of the R&D that is patent does not generate significant return for the firms.

The other important findings are about the average monetary value of patent in different technology. We find that instruments, mechanical and electrical patents are more valuable compared to those in the chemical and pharmaceutical sectors in India. To some extent, this particular result contradicts the common understanding on patent valuation conducted in developed economies. The result implies that some technologies are more valuable in terms of domestic market demand. In India, before 2005 amendments product patents were not allowed, therefore value of process patents are lesser in comparison to other technology. (. Thus, from the policy standpoint this result is important to understand the differential R&D preferences and the nature of market. Other findings of this study reflect that the distribution of patent value is highly asymmetric across technology and ownership group. Large numbers of patents are actually less valuable and only few patents hold high value. The study also finds that the mean patent value increases with an additional renewal year along with other patent characteristics.

---

[i] We do not focus on social value stemming from the patented invention or the value of underlying technology as its calculation requires different set of methodological tools. Social value of a patent includes both future technological developments and the value of current commercial applications under the umbrella of social welfare (Baron and Delcamp, 2011). As opposed to social value, private value of patent only gives the estimate for its owner.

[ii] We provide more discussion on this aspect in the next section.

[iii] Discount rate is fixed at 10% to make comparison with previous studies.

[iv] The detailed reasons and literature support for the direction of variables are given in section 4.2.

[v] http://ipindia.nic.in/writereaddata/Portal/ev/rules/pr80.html